\documentclass[reprint,superscriptaddress,showpacs,preprintnumbers,amsmath,amssymb,aps,prl,floatfix,lengthcheck]{revtex4-1}

\usepackage{graphicx}% Include figure files
\usepackage{dcolumn}% Align table columns on decimal point
\usepackage{bm}% bold math
\usepackage{hyperref}% add hypertext capabilities
\begin{document}

%\preprint{APS/123-QED}

\title{Far-Infrared Conductivity Measurements of Pair Breaking in Superconducting Nb$_{0.5}$Ti$_{0.5}$N Thin Films Induced by an External Magnetic Field }

\author{Xiaoxiang Xi}
\affiliation{Department of Physics, University of Florida, Gainesville, Florida 32611, USA}
\author{J. Hwang}
\affiliation{Department of Physics, University of Florida, Gainesville, Florida 32611, USA}
\affiliation{Department of Physics, Pusan National University, Busan 609-735, Republic of Korea}
\author{C. Martin}
\affiliation{Department of Physics, University of Florida, Gainesville, Florida 32611, USA}
\author{D. B. Tanner}
\affiliation{Department of Physics, University of Florida, Gainesville, Florida 32611, USA}
\author{G. L. Carr}
\affiliation{National Synchrotron Light Source, Brookhaven National Laboratory, Upton, New York 11973, USA}
\date{\today}

\begin{abstract}
We report the complex optical conductivity of a superconducting thin-film of Nb$_{0.5}$Ti$_{0.5}$N  in an external magnetic field. The field was applied parallel to the film surface and the conductivity extracted from far-infrared transmission and reflection measurements. The real part shows the superconducting gap, which we observe to be suppressed by the applied magnetic field. We compare our results with the pair-breaking theory of Abrikosov and Gor'kov and confirm directly the theory's validity for the optical conductivity.
\end{abstract}
\maketitle

Magnetic fields have dramatic effects on the superconducting state; when they are stronger than the upper critical field, superconductivity is destroyed. Fields below this critical value induce supercurrents and also act on the spin and orbital motion of quasiparticle states. An applied field lifts the spin-degeneracy of each electronic state, potentially causing a paramagnetic shift of quasiparticle density of states \cite{PhysRevLett.25.1270} which would give a linear shift of the spectroscopic gap with field \cite{PhysRevB.34.1582}. The effect is noticeable only for materials with a very small spin-orbit scattering rate, in which the spin is a ``good'' quantum number. The field also alters the orbitals of single-particle states from which the BCS ground state is formed, breaking the time-reversal symmetry of the condensate pairing. The result is a finite lifetime for a given Cooper pair and an overall weakening of the superconducting state. This weakening is directly revealed by a reduction in the single-particle gap and forms the basis of the pair-breaking theory originally proposed by Abrikosov and Gor'kov \cite{AG1} to describe the effect of magnetic impurities on superconductivity.  The depairing phenomena can be characterized by a single pair-breaking parameter, $\Gamma$, that depends on whether the theory describes external magnetic fields, supercurrents, spin exchange, or other effects. Maki \cite{PTP.29.603} showed that a thin superconductor in the dirty limit will exhibit pair breaking, equivalent to that caused by magnetic impurities, when subjected to a homogeneous magnetic field. Because paramagnetism and pair breaking can both affect the spectroscopic gap, experimental verification of the pair-breaking effect is simplified in materials with large spin-orbit-scattering  \cite{PhysRevB.8.3161}.

Optical spectroscopy probes directly a superconductor's excitation spectrum, making it ideal for studying the gap evolution under an applied magnetic field. Ordinary metallic superconductors have a gap in their optical spectrum \cite{PhysRev.111.412}, requiring a minimum of $2\Delta$ of photon energy to break Cooper pairs. The gap, which in weak coupling BCS theory is $2\Delta \simeq 3.5kT_c$, makes the $T=0$ real part of the optical conductivity be zero for photon energies below the gap. The missing spectral weight in $\sigma_1(\omega)$ appears as a delta function at zero frequency \cite{PhysRevLett.2.331}. By the Kramers-Kronig relations, the delta function gives a dominant $1/\omega$ form to $\sigma_2(\omega)$ and is responsible for the frequency independent penetration depth, $\lambda_L = c/\sqrt{4\pi\omega\sigma_2}$. This behavior is observed in most metallic superconductors (Sn, In, Pb, Hg, etc.) although strong-coupling effects are sometimes necessary for quantitative 
agreement \cite{PhysRev.104.844}. By determining both the real and imaginary parts of the optical conductivity under applied magnetic field, one can test theories for the magnetic field suppression of the gap. 

We find it somewhat surprising that magnetic-field-induced pair-breaking effects have not been convincingly verified by optical studies. Such effects have been observed in tunneling spectra \cite{PhysRevLett.90.127001} and are hinted at by absorption data \cite{PhysRevB.34.1582}. In addition, the effect of magnetic impurities have been studied in detail \cite{PhysRev.181.774}. In this Letter we report far-infrared transmission and reflection spectra of Nb$_{0.5}$Ti$_{0.5}$N under external magnetic field, applied parallel to the film surface. The extracted optical conductivity $\sigma_1$ demonstrates a suppression of the gap by the field, in quantitative agreement with pair-breaking theory. This is the first time that optical absorption has been employed to  test quantitatively the theory of pair-breaking by an external magnetic field.

\begin{figure}[t]
\includegraphics[width=0.40\textwidth]{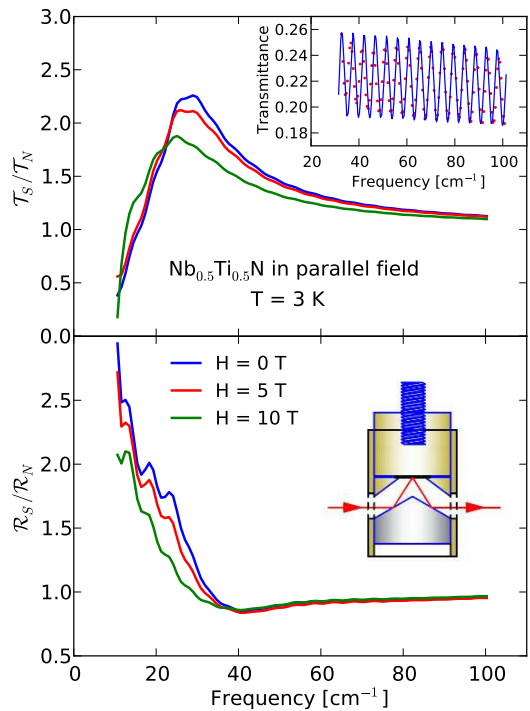}                
\caption{Measured superconducting to normal state ratios for the transmission (upper panel) and reflection (lower panel) at magnetic fields of 0, 5, and 10~T. The weak oscillatory features are partially-resolved multiple internal reflections in the substrate. The inset in the upper panel is the transmittance for $T=300$~K, measured
at 3$\times$ higher spectral resolution. The blue curve is a fit using the optical constants of quartz \cite{Appl.Opt.12.398}. The inset in the lower panel shows the sample holder for reflection measurements. The sample is the thin black slab near the middle. A polished aluminum roof-type mirror reflects the input beam onto the sample at a $30^{\circ}$ angle and then redirects the reflected beam back onto the original optical path. The holder has other interior surfaces angled to minimize stray reflections.} 
\label{TsRs}
\end{figure}

We started with a set of NbTiN thin films of varying thicknesses and substrate materials, and selected one on quartz having $R_{\square}>100~\Omega/\square$ as well as $T_c>10$~K for our magnetic field study. Transmittance data (inset in the upper panel of Fig.~\ref{TsRs}) give the normal-state sheet resistance: $R_{\square} = 146$ $\Omega/\square$. Magnetic susceptibility measurements with a SQUID magnetometer determine $T_c\approx 12$ K, $H_{c1}$ to be between 0.01~T and 0.03~T and $H^{\parallel}_{c2}\approx 15$~T. The optical gap for $T=2$ K ($\ll T_c$) and zero field is 28.5~cm$^{-1}$. The quartz substrate has negligible absorption in the spectral range of interest (10--110 cm$^{-1}$). 

Infrared transmission and reflection measurements were performed at Beamline U4IR of the National Synchrotron Light
Source, Brookhaven National Laboratory; the   high-brightness of the broadband synchrotron radiation is an excellent spectroscopic source. The samples were mounted in a $^4$He Oxford cryostat equipped with a 10 Tesla superconducting magnet; the base sample temperature is 1.6 K. The spectra were collected using a Bruker IFS 66-v/S spectrometer and a high sensitivity, large area B-doped Si composite bolometer operating at 1.8 K; cooled filters limited the upper frequency to 110 cm$^{-1}$.

A sketch of the reflection stage is shown in the inset in the lower panel of Fig.~\ref{TsRs}. The angle of incidence for the reflection measurements is about 30$^{\circ}$; for transmission it is near normal. The magnetic field is applied parallel to the film surface. The field direction is important when considering the behavior of these type II superconductors. For normal field, vortices appear above $H_{c1}$ and form a dense lattice of lossy core material as it approaches $H_{c2}$. We avoided this vortex regime by orienting the field parallel to the film surface. In this case, because the thickness is much smaller than the penetration depth and somewhat smaller than the coherence length, a significant density of vortices is unlikely. The critical field increases, but the order parameter is driven to zero at this larger $H_c$. Nb$_{1-x}$Ti$_x$N typically has a penetration depth $\lambda\approx 100$~nm and a coherence length $\xi\approx 20$~nm \cite{IEEE.Appl.SC.12.1795}, satisfying $\kappa\equiv \lambda/\xi\gg 1$. When a magnetic field is applied parallel to the thin film surface, the vortex spacing is greater than $\xi$ \cite{Parks}, which itself is close to the film thickness. Therefore we do not expect vortex-induced effects to be significant. Moreover, the field decays according to the much larger penetration depth, making the average field in the film be approximately 0.999 of the applied field \cite{IntroSC_Tinkham}. Hence, the field is nearly uniform in the thin film.

Our goal is to extract the optical conductivity of the thin film superconductor from reflection and transmission measurements. Beginning with the pioneering work of Palmer and Tinkham \cite{PhysRev.104.844}, this approach has been used a number of times to study both metallic and cuprate superconductors \cite{PhysRevB.43.10383}. In a conventional transmittance or reflectance measurement, one measures separately the sample and a reference having known optical properties---typically an open aperture with no sample for transmittance and a known metal for reflectance. Sample exchange can lead to errors, especially for the absolute reflection, where sample orientation is critical. To avoid sample exchange errors, we used the sample in the normal state, and at $H=0$~T, for our reference. Specifically, we measured the sample spectrum (transmission or reflection) at different fields ranging from 0 to 10~T in the superconducting state ($T = 3$~K), using the normal state ($T=20$~K), zero-field spectrum for the reference. If required, the relative measurements can be made absolute by measuring the normal-state transmittance and reflectance or by calculating them from the Drude model in the limit $\omega\ll 1/\tau$ (a very good assumption for our films). The directly acquired data are therefore the ratios of transmittance  $\mathcal{T}_S/\mathcal{T}_N$ and reflectance $\mathcal{R}_S/\mathcal{R}_N$, where the subscripts $S$ and $N$ denote superconducting state and normal state respectively. Fig.~\ref{TsRs} shows the data at zero, intermediate and high fields, measured with a resolution of 3.5 cm$^{-1}$, which does not fully resolve the interference fringes from multiple reflections inside the substrate. The peak in $\mathcal{T}_S/\mathcal{T}_N$ shifts to lower frequency as field increases, suggesting the suppression of the energy gap due to the field. The reflection data were corrected for the measured stray light and for the 30$^{\circ}$ angle of incidence (see Supplemental Material) before calculating the optical conductivity. 

The analysis for the thin film optical conductivity $\sigma = \sigma_1 + i\sigma_2$ begins with the expressions for the normal-incidence transmission through, and reflection from, the front film surface of the sample \cite{PhysRev.104.844},
\begin{eqnarray}
\mathcal{T}_f &=& \frac{4n}{\left(Z_0\sigma_1d+n+1\right)^2+\left(Z_0\sigma_2d\right)^2},\label{Tf}\\
\mathcal{R}_f &=& \frac{\left(Z_0\sigma_1d+n-1\right)^2+\left(Z_0\sigma_2d\right)^2}{\left(Z_0\sigma_1d+n+1\right)^2+\left(Z_0\sigma_2d\right)^2},
\label{Rf}
\end{eqnarray}
where $Z_0 \approx 377$~$\Omega$ is the vacuum impedance, $d$ is the film thickness, and $\sigma_1$, $\sigma_2$ are the real and imaginary parts of the optical conductivity of the thin film, either in the superconducting state or in the normal state. In practice, we measure the combination of film and substrate, giving the external transmittance $\mathcal{T}_{ext}$ and external reflectance $\mathcal{R}_{ext}$. If the substrate surfaces are parallel on the scale of the wavelength and the measurement resolution is high enough, these quantities typically show fringes due to partially coherent multiple internal reflections inside the substrate. Smoothing high resolution data or taking measurements with a low resolution produces the incoherent spectrum, where one may add intensities rather than amplitudes. In this case, $\mathcal{T}_{ext} = \mathcal{T}_f (1-\mathcal{R}_Q)e^{-\alpha x}/(1-\mathcal{R}_Q\mathcal{R}_f^{\prime} e^{-2\alpha x})$, where $\mathcal{R}_Q \approx (n-1)^2/(n+1)^2$ is the reflectance of the quartz surface, $\alpha$ is the (small) absorption coefficient of the quartz, $x$ is the thickness of the quartz and $\mathcal{R}_f^{\prime}$ is the film reflection from inside the substrate. There is a similar equation for $\mathcal{R}_{ext}$. Quartz has negligible absorption and dispersion over the spectral and temperature range of interest. Thus we take $\alpha=0$ and $n=2.12$ (a constant), yielding $\mathcal{R}_Q \approx 0.13$.

Our measurements give us the external transmission and reflection ratios, $\mathcal{T}_{ext,S}/\mathcal{T}_{ext,N}$, and $\mathcal{R}_{ext,S}/\mathcal{R}_{ext,N}$ that include the substrate. For range of conductivity values expected for the film, we find that, to a very good approximation, $\mathcal{T}_{ext,S}/\mathcal{T}_{ext,N} = \mathcal{T}_S/\mathcal{T}_N$ and $\mathcal{R}_{ext,S}/\mathcal{R}_{ext,N} = \mathcal{R}_S/\mathcal{R}_N$. The normal-state transmittance and reflectance can be derived from Eqs.~\eqref{Tf} and~\eqref{Rf} by setting $\sigma_1 = \sigma_N$ and $\sigma_2 = 0$, $\mathcal{T}_N = 4n/(Z_0 \sigma_N d + n + 1)^2,\mathcal{R}_N = (Z_0 \sigma_N d+n-1)^2/(Z_0 \sigma_N d + n + 1)^2$. Here $\sigma_N$ is related to the normal state sheet resistance of the thin film $R_{\square} = 1/\sigma_Nd$, which we have determined from the normal-state transmittance. Hence we know $\mathcal{T}_N$ and $\mathcal{R}_N$ and may use them to calculate $\mathcal{T}_S$ and $\mathcal{R}_S$ from our measured ratios. Then we invert Eqs.~\eqref{Tf} and~\eqref{Rf} to find 
\begin{eqnarray}
\frac{\sigma_1}{\sigma_N} &=& \frac{nR_{\square}}{Z_0}\frac{1-\mathcal{R}_S-\mathcal{T}_S}{\mathcal{T}_S},
\label{sigma1}\\
\frac{\sigma_2}{\sigma_N} &=&
\frac{R_{\square}}{Z_0}\left[\frac{4n}{\mathcal{T}_S}-\left(Z_0\sigma_1d+n+1\right)^2\right]^{1/2}.
\label{sigma2}
\end{eqnarray}

\begin{figure}[t]
\includegraphics[width=0.465\textwidth]{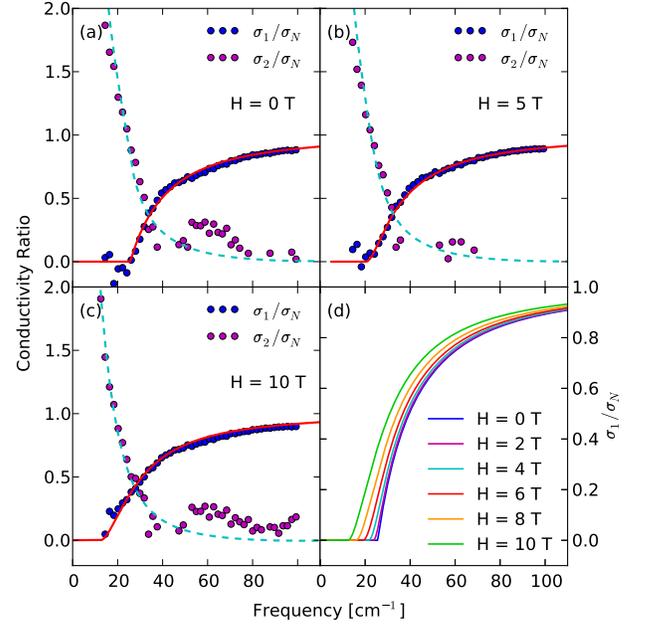}                
\caption{(a)--(c) The real and imaginary parts of the $T=3$~K superconducting state optical conductivity (normalized to the normal state conductivity) at three different applied magnetic fields. The solid lines are fits to $\sigma_1/\sigma_N$ using the pair-breaking theory. The dashed lines show the corresponding $\sigma_2/\sigma_N$ as determined by a Kramers-Kronig transform of the real part. (d) The fitted $\sigma_1/\sigma_N$ at six fields.} 
\label{s1s2}
\end{figure}

The normalized optical conductivity at 0, 5, and 10~T are shown in panels (a)--(c) in Fig.~\ref{s1s2}. 
$\sigma_2/\sigma_N$ has some data points missing because the term under the square root in Eq.~\eqref{sigma2} is not guaranteed to be positive for the measured transmission and reflection when noise is included. A weak interference fringe in both the transmission and reflection measurements results in the excess $\sigma_2/\sigma_N$ over the 40 to 80~cm$^{-1}$ range. The solid lines are fits to the data using the pair-breaking theory as extended by Skalski \textit{et al.} \cite{PhysRev.136.A1500} to calculate $\sigma_1/\sigma_N$ at 0~K:
\begin{eqnarray}
\frac{\sigma_1}{\sigma_N} = \frac{1}{\omega}\int^{-\Omega_G + \omega/2}_{\Omega_G - \omega /2}dq[n(q+\omega /2)n(q-\omega /2)\nonumber\\
+m(q+\omega /2)m(q-\omega /2)]
\label{s1sn_eqn}                          
\end{eqnarray}
for $\omega\ge 2\Omega_G$ and zero otherwise, where $n(q) = \mathrm{Re}(u/\sqrt{u^2-1})$ and $m(q) = \mathrm{Re}(1/\sqrt{u^2-1})$. $u$ is the solution to $u\Delta = q + i\Gamma u/\sqrt{u^2-1}$, with $\Delta$ the pair-correlation gap. This, in turn, can be determined from the pair-breaking parameter $\Gamma$ and the zero field excitation gap $\Delta_0$ using $\ln(\Delta/\Delta_0)=-\pi\Gamma/4\Delta$ for $\Gamma<\Delta$. $\Omega_G$ in the integration limits is the effective spectroscopic gap, $\Omega_G = \Delta[1-(\Gamma/\Delta)^{2/3}]^{3/2}$ for $\Gamma<\Delta$. We fit our $H = 0$~T results to determine $\Delta_0$, and then proceeded to fit $\sigma_1/\sigma_N$ for $H > 0$~T using only $\Gamma$ as an adjustable parameter. The imaginary part of the conductivity (dashed lines) was calculated by a Kramers-Kronig transform of the real part. The temperature $T \approx 0.25 T_c $ is low enough that the gap has reached its zero-temperature value. The zero-field case reduces to the standard BCS Mattis-Bardeen \cite{PhysRev.111.412} description of a dirty-limit superconductor. Panel (d) in Fig.~\ref{s1s2} shows the fitted $\sigma_1/\sigma_N$ at six different fields. Clearly, the absorption edge moves to lower energy as the field increases. The field-induced pair breaking also smears out the gap-edge singularity in the quasiparticle density of states \cite{PhysRev.136.A1500}, so that the initial rise of $\sigma_1$ becomes less abrupt for increasing fields. This slower increase is evident when comparing the 0~T and 10~T results in Fig.~\ref{s1s2}.  

 The oscillator-strength sum rule $\int^{\infty}_0\sigma_1(\omega)d\omega=\pi ne^2/2m$, where $n$ is the electron density and $e$ and $m$ are the charge and mass of the electron, requires the area under $\sigma_1(\omega)$ to be the same for normal and superconducting states. The ratio $\sigma_1/\sigma_N$ in Fig.~\ref{s1s2} is always less than unity; the ``missing area'' condenses to a $\delta$-function at zero frequency that is a measure of pair condensate density and is directly related to the pair-correlation gap. Panel (d) in Fig.~\ref{s1s2} therefore shows a weakening of superconductivity as field increases. There is a limit in which the absorption edge approaches 0, while the missing area remains finite. The superconductor enters a ``gapless'' region but still maintains superconducting properties. 

\begin{figure}[h]
\includegraphics[width=0.40\textwidth]{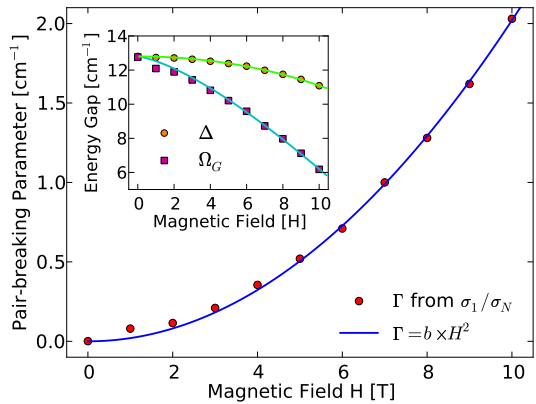}                
\caption{Field dependence of the pair-breaking parameter $\Gamma$, determined from the experimental optical conductivity (circles) along with a fit to $\Gamma=bH^2$ (solid line). The inset shows the pair correlation gap $\Delta$ and the effective spectroscopic gap $\Omega_G$ for the same fields. The solid lines are theoretical predictions using the pair-breaking theory and the fitted value of $b$.} 
\label{GammaH}
\end{figure}

The quantity $\Gamma$ describes the strength of pair-breaking and, for the range of fields used here, is quadratic in field,  $\Gamma=bH^2=\tau_{\mathrm{tr}}v_f^2(eHd)^2/18$, where $\tau_{\mathrm{tr}}$ is the transport collision time and $d$ is the film thickness \cite{Parks}. $\Gamma$, as extracted from our data at different fields, is plotted in Fig.~\ref{GammaH}. 
The quadratic fit is good, yielding $b = 0.020$ cm$^{-1}$/T$^2$. We estimate $\tau_{\mathrm{tr}}$ from $\sigma_N = ne^2\tau_{\mathrm{tr}}/m$ and $R_{\square}=1/\sigma_N d$, $\tau_{\mathrm{tr}}=m/R_{\square}dne^2\approx 1.51\times 10^{-16}$ s, where $R_{\square} = 146$~$\Omega/\square$, $d=10$~nm and $n\approx 1.61\times 10^{23}$~cm$^{-3}$ is the electron density of NbN \cite{PhysRevB.77.214503} similar to that of NbTiN. If we take the Fermi velocity to be that of NbN \cite{PhysRevB.77.214503}, $v_f\approx 1.95\times 10^8$~cm/s, then, $b = 0.039$~cm$^{-1}/$T$^2$, consistent with the fitted value within the uncertainty of the materials parameters.  

The shift of the excitation energy gap $2\Omega_{G}$ due to the application of magnetic field has already been discussed, and can be seen from the absorption edge in $\sigma_1/\sigma_N$. This gap $\Omega_G$ and the pair-correlation gap $\Delta$ are calculated and compared in the inset of Fig.~\ref{GammaH}. Both $\Omega_G$ and $\Delta$ drop as field increases, but the reduction of $\Omega_G$ is much greater at any given field. The sample at the highest attainable field of 10~T is still far away from the gapless region where $\Omega_G$ vanishes. The experimental and theoretical values of $\Omega_G$ and the pair-correlation gap $\Delta$ are in excellent agreement, as shown in the inset of Fig.~\ref{GammaH}.

In conclusion, we measured far-infrared transmission and reflection of a thin-film superconductor in a magnetic field parallel to the film surface. The real and imaginary parts of the optical conductivity are derived from these data, the former showing the absorption edge depressed due to the applied in-plane field. The degree of suppression is in good agreement with the pair-breaking theory. 

This work was supported by the U.S. Department of Energy through contract No. DE-ACO2-98CH10886 at the NSLS and Contract 
No.~DE-FG02-02ER45984 at the University of Florida. J. Hwang acknowledges financial support from the National Research Foundation of Korea (No. 20090074977). We are grateful to P. J. Hirschfeld for valuable discussions and to G. Nintzel for technical support. We thank P. Bosland and E. Jacques (CEA-Saclay) for providing the NbTiN samples.

%\bibliographystyle{apsrev4-1}
%\bibliography{reference}% Produces the bibliography via BibTeX.

\end{document}